\documentclass[prd,12pt,aps,superscriptaddress,preprintnumbers,showpacs,tightenlines,floatfix,nofootinbib,amssymb]{revtex4}
\usepackage{epsfig}

\newcommand{\beqn}{\begin{eqnarray}}
\newcommand{\eeqn}{\end{eqnarray}}
\newcommand{\eq}[1]{(\ref{#1})}
\newcommand{\cC}{{\cal C}}
\newcommand{\cY}{{\cal Y}}
\newcommand{\cP}{{\cal P}}
\newcommand{\dd}{{\mathrm d}}
\newcommand{\Z}{{Z \!\!\! Z}}
\def\bbbone{{\mathchoice {\rm 1\mskip-4mu l} {\rm 1\mskip-4mu l}
{\rm 1\mskip-4.5mu l} {\rm 1\mskip-5mu l}}}
\newcommand{\Kanazawa}{\affiliation{Institute for Theoretical Physics,
Kanazawa University, Kanazawa 920-1192, Japan}}
\newcommand{\ITEP}{\affiliation{Institute of Theoretical and
Experimental Physics, B.Cheremushkinskaya 25, Moscow, 117259, Russia}}

\begin{document}

\title{A gauge-invariant object in non-Abelian gauge theory}

\author{M.N.~Chernodub}\ITEP\Kanazawa

\preprint{ITEP-LAT/2005-06}
\preprint{KANAZAWA/2005-04}

\begin{abstract}
We propose a nonlocal definition of a gauge-invariant object in
terms of the Wilson loop operator in a non--Abelian gauge theory.
The trajectory is a closed curve defined by an (untraced) Wilson
loop which takes its value in the center of the color group. We
show that definition shares basic features with the
gauge-dependent 't~Hooft construction of Abelian monopoles in
Yang-Mills theories. The chromoelectric components of the gluon
field have a hedgehog-like behavior in the vicinity of the object.
This feature is dual to the structure of the 't~Hooft-Polyakov
monopoles which possesses a hedgehog in the magnetic sector. A
relation to color confinement and lattice implementation of the
proposed construction are discussed.
\end{abstract}


\date{March 14, 2005; Revised: December 22, 2005}

\maketitle

\section{Introduction}
\label{one}

The mechanism of color confinement in QCD is an important problem
which is not yet solved. An approach by Nambu, 't~Hooft and
Mandelstam~\cite{ref:DualSuperconductor} suggests that the vacuum
of QCD can be treated as a dual superconductor, which confines
quarks due to presence of special configurations of gluon fields
called ``Abelian monopoles''. In brief, if the monopoles are
condensed then the chromoelectric flux of (anti-) quarks is
squeezed into tubes (``QCD strings'') which confine quarks and
anti-quarks into tightly bound colorless bound states. Features of
this mechanism and results of corresponding numerical studies on
the lattice -- which confirm the validity of the dual
superconductor mechanism in a particular gauge of the Yang-Mills
theory -- can be found in reviews~\cite{ref:Reviews}.

The basic element of the dual superconductor approach is the
Abelian monopole. The existence of this object is not supported by
the symmetries of QCD. However, the monopoles can be identified
with particular configurations of the gluon fields by the
so-called Abelian projection formalism invented by
't~Hooft~\cite{ref:thooft:monopole}. This formalism relies on a
partial gauge fixing of the $SU(N)$ gauge symmetry up to an
Abelian subgroup. The Abelian monopoles appear naturally in the
Abelian gauge as a result of the compactness of the residual
Abelian group.

One of the major problems associated with the 't~Hooft
construction of the monopole in pure gauge theories is that this
construction is not universal. Or, as often said, the construction
is gauge dependent: the location of the monopole in a fixed gluon
field configuration is dependent of the gauge fixing matrix $X$
(to be defined in the next Section). There is an infinite number
of such matrices, and, respectively, there is an infinite number of
the Abelian monopole definitions in non-Abelian gauge theories.
Some (if not most) of the definitions are physically irrelevant
for infrared problems such as the confinement problem. For
example, the Abelian monopoles existing in the Polyakov Abelian
gauge -- which is defined by diagonalization of the Polyakov loop
-- are always static in the continuum
limit~\cite{ref:Yukawa,ref:Polyakov:gauge}, and they cannot
contribute to the confinement of the static quarks. On the other
hand, the monopoles defined in the so-called Maximal Abelian
gauge~\cite{ref:MA} were numerically shown to make a dominant
contribution to the confinement of
quarks~\cite{ref:monopole:dominance} as well as to other
low-energy non-perturbative phenomena.

Clearly, the gauge invariant phenomena (such as the quark
confinement) can not be described by the gauge-dependent
mechanism. Probably, one should blame the tool of Abelian
projections which is used to "detect" the monopoles: the monopoles
(as confining defects) are observed well in one gauge and they
are ``eaten up'' or contaminated by artifacts in another gauge.

There were various attempts to find an appropriate solution of the
gauge-dependence problem of the monopole-based confinement mechanism.
Some approaches~\cite{ref:continuum,ref:Fedor,ref:Faber} are based on an attempt
to find a gauge invariant definition of Abelian monopoles in non-Abelian gauge
theories. Another approach~\cite{ref:Suzuki:Landau} pays attention
to the magnetic displacement currents observed in the Landau
gauge. The displacement is linked to the existence $A^2$
condensate~\cite{ref:A2} which should contain a gauge--invariant
piece. On the other hand the authors of Ref.~\cite{ref:Alex} claim
that in a pure non-Abelian gauge theory the monopole charge can
not be defined at all. This claim is opposed by Ref.~\cite{ref:reply:Alex}.

In this paper we propose a construction of a new gauge-invariant
object in non-Abelian gauge theories, which may have a tight link
to the confinement phenomena. The basic idea is to use the Wilson
loop variable as an effective path-dependent Higgs field which
allows to define a singularity in self-consistent way. The
construction resembles the 't~Hooft definition of the Abelian
monopole but does not rely on any gauge fixing procedure.

The structure of the paper is as follows. In
Section~\ref{sec:tHooft} we briefly overview the Abelian monopoles
defined in the Abelian projection formalism. In
Section~\ref{ref:construction} we show how the gauge-invariant
object can uniquely be constructed in a manner similar to the
Abelian monopoles. A relation to the confinement of color and a
possible lattice implementation of the proposed construction are
also discussed. In Section~\ref{sec:examples} we consider examples
of specific gluon field configurations and show that our
definition correctly recognizes a self-dual BPS monopole. We also
discuss dynamics of our objects at a finite-temperature as well as
in lower dimensions. The last Section contains our conclusion.

\section{'t~Hooft's Abelian monopoles in SU(N) gauge theories}
\label{sec:tHooft}

The 't~Hooft definition of an Abelian monopole~\cite{ref:thooft:monopole} in non--Abelian
gauge theories is based on a partial gauge fixing of the non--Abelian gauge freedom up to
the Abelian one. Technically, the Abelian gauge fixing is achieved by diagonalization\footnote{Note that
there is also a different class of the Abelian gauges which can not be defined by diagonalization of any operator.
The Maximal Abelian gauge~\cite{ref:MA} -- which is most popular nowadays -- belongs to this class.} of an arbitrary
{\it adjoint} operator $X(x) \equiv X^a T^a$, where $T^a$ are the generators of the $SU(2)$
group, $a = 1,\dots, N^2-1$. In the Abelian gauge the operator $X$ is diagonalized by the gauge transformations
in the whole space-time:
\beqn
X(x) \to X'(x) \equiv \Omega^\dagger(x) \, X(x) \, \Omega(x)
= {\mathrm{diag}}\Bigl(\lambda_1(x),\lambda_2(x),\dots, \lambda_N(x)\Bigr)
\equiv X_{\mathrm{diag}}(x)\,,
\label{eq:Xdiag}
\eeqn
where $\Omega$ is the matrix of the $SU(N)$ gauge transformation and $\lambda_1 \geqslant \lambda_2 \geqslant \dots \geqslant \lambda_N$.

{}From Eq.~\eq{eq:Xdiag} one sees that the gauge can not be fixed completely since the diagonal operator
$X_{\mathrm{diag}}$ remains intact under arbitrary {\it Abelian} ${[U(1)]}^{N-1}$ gauge transformations given by
matrices
\beqn
\Omega^{\mathrm{Abel}} = {\mathrm{diag}}\Bigl(e^{i \alpha_1(x)},e^{i \alpha_2(x)},\dots,e^{i \alpha_N(x)}\Bigr)\,,
\qquad \sum^N_{a=1} \alpha_a(x) = 1\,,
\label{eq:Omega:Abel}
\eeqn
where $\alpha_i(x)\in [0,2\pi)$ are arbitrary functions which obey the $SU(N)$--imposed constraint.

After the $SU(N) \to {[U(1)]}^{N-1}$ gauge fixing~\eq{eq:Xdiag}
the model not only respects the Abelian gauge invariance but it
also possesses the topological defects called Abelian
monopoles~\cite{ref:thooft:monopole}.
The Abelian monopoles come from singularities of the gauge fixing
condition~\eq{eq:Xdiag}. If at the point $x_0$ two eigenvalues of
the diagonalized $X$-matrix coincide (say,
$\lambda_i(x_0)=\lambda_{i+1}(x_0)$) then at this point the
Abelian gauge can not be fixed by the condition~\eq{eq:Xdiag}. In
other words, in the point $x_0$ the matrix of the residual gauge
transformations is no more diagonal contrary to the Abelian
matrix~\eq{eq:Omega:Abel}. Formally, the residual gauge
transformations also contain the $SU(2)$ subgroup embedded into
the $SU(N)$ group at the place where the columns with the numbers
$i$ and $i+1$ overlap with the rows $i$ and $i+1$.

Below we consider the $SU(2)$ group. The generalization to the $SU(N)$ case is simply given by repeating
of the $SU(2)$ considerations with respect to the $i$ and $i+1$ columns/rows of the $SU(N)$ $X$-matrix
with coinciding eigenvalues $\lambda_i(x_0)=\lambda_{i+1}(x_0)$. In the $SU(2)$ case, $X=X^a \sigma^a/2$
where $\sigma^a$, $a=1,2,3$ are the Pauli matrices,
and the diagonal matrix is $X_{\mathrm{diag}} = {\mathrm{diag}}\bigl(\lambda,-\lambda\bigr)$ with $\lambda>0$.
The gauge fixing singularities appear at points of the space-time where the $X$-matrix is degenerate,
\beqn
X(x)=0\,,
\label{eq:X.eq.0}
\eeqn
{\it i.e.} where $\lambda(x)=0$. Since $\lambda(x) =
(1/2){\bigr(\sum^3_{a=1}X^{a,2}\bigl)}^{1/2}$, the single condition
$\lambda(x)=0$ is equivalent to the three {\it independent}
constraints $X^a(x)=0$, $a=1,2,3$. These three constraints in
four-dimensional space-time define a loop (or, a set
of loops). This loop is the world trajectory of the Abelian
monopole. One can show that the monopole charge is conserved~\cite{ref:thooft:monopole,ref:general:singularities}.

Consider the point $x_0$ at the monopole trajectory. Without loosing of generality let us assume that the monopole current
is pointing out towards 4th direction. Then the matrix $X$ has the following spatial structure in the vicinity of the monopole:
\beqn
X(x) = \frac{\sigma^a}{2} \, Y^{ai} {\bigl(x - x_0\bigr)}^i + O\Bigl(\bigl(x - x_0\bigr)^2\Bigr)\,,\qquad
Y^{ai} \equiv \frac{\partial X^a(x)}{\partial x^i}{\Biggl|}_{x=x_0}\,, \qquad a,i=1,2,3.
\label{eq:X:prelim}
\eeqn
If the $Y$--matrix is not degenerate, ${\mathrm{det}} Y \neq 0$,
then the point $x_0$ corresponds to the isolated singularity point
(which is the general case to be considered below).

Using the change of variables in Eq.~\eq{eq:X:prelim}, $x_i \to y_i = {(Y^{-1})}_{ia} (x - x_0)^a$, we transform the $X$-matrix
to the canonical form: $X = (\sigma^a/2)\, y_a +O(y^2)$. Thus, in the vicinity of the singularity, the field $X$ has a
hedgehog form resembling the behavior of the adjoint Higgs field in the vicinity of the 't~Hooft-Polyakov
monopole solution~\cite{ref:tHooft:GGmonopole,ref:Polyakov:GGmonopole} of the Georgi--Glashow model. In order to figure out that the singularity
in the Abelian gauge corresponds indeed to an Abelian monopole, one notes that a gauge transformation -- which diagonalizes
the singular hedgehog configuration into the regular (diagonal) configuration -- must itself be singular.
Using the standard parameterization, ${\mathbf y} = r\,(\sin \theta, \cos \varphi,\sin \theta, \sin \varphi, \cos \theta)$,
one can write the matrix $X$ and the diagonalization matrix $\Omega$, Eq.~\eq{eq:Xdiag}, as follows
(we omit $O(y^2)$ corrections below):
\beqn
X =\frac{r}{2} \left(
\begin{array}{lr}
\cos \theta & e^{-i \varphi}\sin \theta \\
e^{i \varphi}\sin \theta & - \cos \theta \\
\end{array}
\right)\,,
\qquad
\Omega = \left(
\begin{array}{lr}
\cos {\theta /2} & e^{- i \varphi} \sin {\theta /2}\\
- e^{i \varphi} \sin {\theta /2} & \sin {\theta /2}\\
\end{array}
\right)
\label{eq:X:Omega}
\eeqn
Then, in the Abelian gauge the field strength tensor for the Abelian field is (here and in the next equation superscript
denotes the color component):
\beqn
f_{\mu\nu} = \partial_{[\mu,} a_{\nu]} \equiv \partial_\mu {\bigl(A^\Omega\bigr)}^3_\nu
- \partial_\nu {\bigl(A^\Omega\bigr)}^3_\mu =
f^{\mathrm{reg}}_{\mu\nu} + f^{\mathrm{sing}}_{\mu\nu}\,,
\label{eq:fmunu:prel}
\eeqn
where $a_\mu \equiv A^\Omega_\mu = \Omega^\dagger \bigl(A_\mu + i/g \cdot \partial_\mu \bigr) \Omega$ is the diagonal
component of the gluon field $A_\mu$ rotated to the Abelian gauge by the gauge transformation $\Omega$, and $g$ is the Yang-Mills gauge coupling.
The regular part of the Abelian field strength tensor~\eq{eq:fmunu:prel}, $f^{\mathrm{reg}}_{\mu\nu}$ contains single derivatives
of the original gluon field $A_\mu$ and of the gauge matrix $\Omega$. The singular part $f^{\mathrm{sing}}_{\mu\nu}$ contains a commutator
of two derivatives,
\beqn
f^{\mathrm{sing}}_{\mu\nu} = \frac{i}{g} {\Bigl[\Omega^\dagger \partial_{[\mu,} \partial_{\nu]} \Omega\Bigr]}^3 =
- 4 \pi \bigl(\delta_{\mu,1}\delta_{\nu,2} - \delta_{\nu,1}\delta_{\mu,2} \bigr)\, \delta(x_1)\, \delta(x_2) \, \Theta(-x_3)\,,
\label{eq:f:sing}
\eeqn
where $\Theta(x)$ is the Heavyside step function. The singular part corresponds~\eq{eq:f:sing} to the static
Abelian Dirac string located in the $(3,4)$-plane. The string is semi-infinite: it starts at spatial infinity,
$x_1=x_2=0$, $x_3 \to -\infty$, and ends on the static monopole located at origin, $x_1=x_2=x_3=0$.
The magnetic current $k_\mu$ is given by the formula:
\beqn
k_\mu(x) = \frac{1}{8\pi}  \, \varepsilon_{\mu\nu\alpha\beta} \, \partial_\nu f_{\alpha\beta}(x)\,.
\label{eq:k:definition}
\eeqn
The only contribution to $k_\mu$ is given by the singular component \eq{eq:f:sing}. The direct evaluation gives, obviously, that the
monopole current is just a boundary of the Dirac string: $k_\mu(x) = \delta_{\mu,4} \delta(x_1) \delta(x_2) \delta(x_3)$.

Note that due to the diagonalization condition the singularity
manifests itself only in the diagonal ({\it i.e.}, 3rd in color)
component of the gauge field. In order to evaluate the
commutator~\eq{eq:f:sing} we have used the explicit form for the
matrix $\Omega$ in Eq.~\eq{eq:X:Omega} and also implicitly assumed
(without loss of generality) that the $Y$-matrix from
Eq.~\eq{eq:X:prelim} is $Y^{ai} \propto \delta^{ai}$. If the
matrix $Y$ is not diagonal than the direction of the Dirac string
is different from the 3rd spatial direction.

The Abelian field-strength tensor~\eq{eq:fmunu:prel} and, consequently,
the Abelian monopole, can equivalently be defined with
the help of the 't~Hooft field-strength tensor~\cite{ref:tHooft:GGmonopole,ref:Arafune},
\beqn
f_{\mu\nu} \equiv \Phi^a F^a_{\mu\nu} - \frac{1}{g} \, \varepsilon^{abc}\, \Phi^a \,
{\bigl[D_\mu(A) \Phi\bigr]}^b \, {\bigl[D_\nu(A) \Phi\bigr]}^c\,,
\label{eq:tHooft}
\eeqn
where $F^a_{\mu\nu} = \partial_{[\mu,} A^a_{\nu]} + e\, \varepsilon^{abc}\, A^b_\mu \, A^c_\nu$ is
the standard non-Abelian field strength tensor,
${\bigl[D_\mu(A) \Phi\bigr]}^a = \partial_\mu \Phi^a + g \, \varepsilon^{abc}\, A^b_\mu \Phi^c$
is the long derivative, and $\Phi^a$ is the unit composite Higgs field of the unit length,
\beqn
\Phi(x) = \Omega^\dagger (x) \, \sigma^3 \, \Omega(x) \,,\qquad \Phi^a \Phi^a =1\,.
\label{eq:Phi}
\eeqn
The Higgs field is made of the gauge rotation matrices $\Omega$ defined by the
gauge fixing condition~\eq{eq:Xdiag}. In the Abelian gauge, where the diagonalization matrix $X$ is diagonal,
the effective Higgs field $\Phi$ automatically becomes fixed to the Unitary gauge, $\Phi^a = \delta^{a,3}$ and
the 't~Hooft tensor explicitly coincides with the Abelian field-strength tensor~\eq{eq:fmunu:prel} with singular
Abelian field. Discussion on the Abelian singularities in pure non-Abelian gauge
fields after the Abelian gauge fixing can also be found in
Refs.~\cite{ref:thooft:monopole,ref:general:singularities,ref:Yukawa,ref:Jahn}.

\section{A definition of a gauge-invariant object}
\label{ref:construction}

The crucial role in identifying of the monopole singularities in
pure non-Abelian gauge theory is played by the matrix $X$,
Eq.~\eq{eq:X:prelim}, which is a function(al) of the gluon fields.
This matrix can be used to construct, in turn, an effective
adjoint Higgs field, $\Phi$, Eq.~\eq{eq:Phi}. Any particular
choice of the matrix $X$ corresponds to a fixing of a particular
Abelian gauge. The matrix $X$ can be chosen in infinitely many
ways and no physically motivated choice exist {\it a priori}. In
this Section we show that the definition of a gauge-invariant
object (in a manner similar to the Abelian monopole definition)
does exist and that the construction does not correspond to a any
particular gauge fixing.

\begin{figure}[!thb]
\begin{center}
\includegraphics[scale=1.,clip=true]{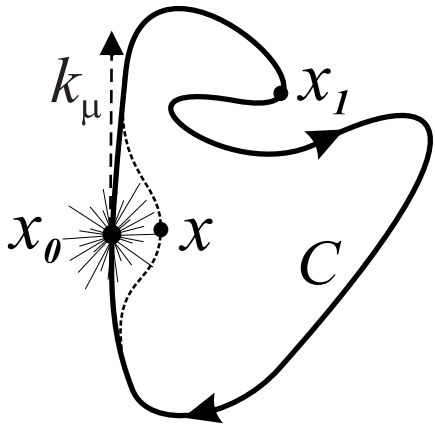}
\end{center}
\vspace{-4mm}
\caption{The contour $\cC$ (solid line) and its variation (dotted line) used in definition of the monopole.}
\label{fig:contour:general}
\end{figure}

Consider a trajectory $\cC$ as depicted in Figure~\ref{fig:contour:general}.
To make a rigorous definition of an object associated with the trajectory $\cC$, let us study an
untraced Wilson loop which starts and ends at some point $x_0 \in \cC$:
\beqn
W_\cC(x_0) = P\,\exp\Bigl\{i g\, \oint_{\cC} \dd x_\mu\, A_\mu(x) \Bigr\}\,.
\label{ref:W}
\eeqn
This operator transforms under the gauge transformations as an adjoint operator, similarly to the operator $X$,
Eq.~\eq{eq:Xdiag}. To make an analogy with $X$ closer, we subtract the singlet part from~\eq{ref:W}, defining
another adjoint operator:
\beqn
\Gamma_\cC(x_0) = W_\cC(x_0) - \bbbone \cdot \frac{1}{2}\, {\mathrm{Tr}}\, W_\cC(x_0)\,.
\label{ref:Gamma}
\eeqn
By the construction, the operator $\Gamma$ is traceless, and under the gauge transformations
it behaves as follows $\Gamma_\cC(x_0) \to \Omega^\dagger(x_0) \Gamma_\cC(x_0) \Omega(x_0)$.

The gauge-invariant condition for our object to have the loop $\cC$ as its world-line trajectory, is
to require that the untraced Wilson loop~\eq{ref:W} belongs to the center of the group, $W_\cC(x_0) \in \Z_2$.
Then the matrix $\Gamma_\cC$ vanishes,
\beqn
\mbox{Condition 1}:\qquad W_\cC(x_0) = \pm \bbbone\qquad \Leftrightarrow\qquad \Gamma_\cC(x_0)=0\,.
\label{eq:condition:1}
\eeqn
This condition is gauge invariant since the eigenvalues of the Wilson loop are gauge invariant quantities.
Equation~\eq{eq:condition:1} is very similar to Eq.~\eq{eq:X.eq.0} which appears in
the 't~Hooft construction. The matrix $\Gamma$ plays the role of the diagonalization matrix $X$.
The condition~\eq{eq:condition:1} is required but not sufficient criterium for our object
to have the loop $\cC$ as its world-line trajectory.

The condition~\eq{eq:condition:1} is self-consistent in a sense, that if the matrix $\Gamma$ vanishes at
the point $x_0$, then it also vanishes at any other point along the trajectory~$\cC$. To show this, let us
consider an arbitrary point $x_1$ at the contour $\cC$, Figure~\ref{fig:contour:general}. Then the Wilson
loops~\eq{ref:W} open at the points $x_0$ and $x_1$ are related to each other by the adjoint transformation:
\beqn
W_\cC(x_1) = U^\dagger(x_1,x_0)\,W_\cC(x_0)\, U(x_1,x_0)\,, \qquad U(x_1,x_0) =
P\,\exp\Bigl\{i g\, \oint^{x_1}_{x_0} \dd x_\mu\, A_\mu(x) \Bigr\}\,.
\eeqn
Obviously, if $W_\cC(x_0) \in \Z_2$ then $W_\cC(x_1) \in \Z_2$ as well, and,
consequently, $\Gamma_\cC(x_1) \equiv 0$. Thus the condition~\eq{eq:condition:1} is
in fact independent on the reference point $x_0$.

In order to show that the simple condition~\eq{eq:condition:1}
does indeed have a similarity with the monopole, let us deform
infinitesimally the contour $\cC \to \cC' = \cC + \delta \cC$ in
the vicinity of the point $x_0$, Figure~\ref{fig:contour:general}.
After the deformation the point $x_0$ is shifted to the new
location $x$ (for the sake of simplicity we do not further use the
prime symbol in $\cC'$). The infinitesimal deformations in the
tangent direction to the current $k_\mu(x_0)$ should not change
value of $\Gamma_\cC$ as we have just seen. Therefore a
non-trivial variation should only be done in the direction,
perpendicular to $k_\mu(x_0)$. Without loss of generality let us
assume that the current is $k_\mu \propto \delta_{\mu,4}$ and then
the infinitesimal vector $x - x_0$ should only have spatial
non-zero components.

In general case the deformed loop $\Gamma_\cC(x)$ should have a
hedgehog-like structure in the vicinity of the zero-point $x_0$,
\beqn
\mbox{Condition 2}:\qquad \Gamma_\cC(x) = \frac{\sigma^a}{2} \, \cY^{ai}_\cC
{\bigl(x - x_0\bigr)}^i + O\Bigl(\bigl(x - x_0\bigr)^2\Bigr)\,,
\label{eq:condition:2}
\eeqn
where $\cY$ matrix is similar to the matrix $Y$ of Eq.~\eq{eq:X:prelim}.
This is the second condition for a hedgehog--like object to be located at
point $x_0$. One may interpret our construction in terms of the Abelian
monopoles which appear in the Abelian gauge formalism. The
diagonalization of the Wilson loop $W_\cC$ corresponds to a
"dynamical Abelian gauge". If the eigenvalues of the loop $W_\cC$
coincide, and the hedgehog structure~\eq{eq:Y:variation} appears,
then we have in this dynamical Abelian gauge the singularity on
the trajectory $\cC$.

Note that Condition~1 follows from Condition~2, while the opposite
is not valid in general. The situation is quite similar to the
Abrikosov vortex solution in the Ginzburg-Landau model: in the
center of the vortex the scalar field is zero (analogously to
Condition~1). However, not all zeros of the scalar fields are
vortices since the singular behavior of the phase of the scalar
field is required as well (an analog of Condition~2). Similarly to
our case, the singular phase of the Higgs field guarantees the
vanishing of the scalar field in the center of the vortex while
the opposite is not generally correct.

As we have seen, Condition~1, Eq.~\eq{eq:condition:1}, is
self-consistent in a sense, that is it fulfilled in any point
$x_0$ of the loop $\cC$ for isolated "singularities". A similar
statement should be valid in general case for Condition~2,
Eq.~\eq{eq:condition:2}. Indeed, as we travel along the loop, the
hedgehog condition~\eq{eq:condition:2} may cease to be valid
provided the matrix $\cY^{ai}_{\cC}$ becomes degenerate at some
point $x_s$ of the loop $\cC$, $\mathrm{det} \cY_\cC (x_s) =0$.
However, the degeneracy means that the singularity is no more
isolated, contradicting the initial assumption.

Note, that formally the matrix $\cY_\cC$ introduced in Eq.~\eq{eq:condition:2} may be defined
as a variation of the loop $\cC$:
\beqn
\cY^{ai}_\cC(x) = \frac{\delta \Gamma_\cC^a(x)}{\delta x^i}{\Biggl|}_{x \to x_0}\,, \qquad a,i=1,2,3\,.
\label{eq:Y:variation}
\eeqn
This equation is very similar to Eq.~\eq{eq:X:prelim} with the
only exception: instead of the usual derivative, in the definition
of the matrix $\cY_\cC$ we have formally used the path derivative
$\delta$ which defines a change of the functional $\Gamma_\cC$
under the infinitesimal change of the contour $\cC$. The variation
itself is infinitesimally small, since the change in the loop
functional $W_\cC$ is proportional to the area of the loop
variation.

In a Lorentz--invariant form the matrix $\cY_\cC$ can be written as follows:
\beqn
\cY^{a\nu}_\cC(x) \propto m^\mu_\cC\, F^a_{\mu \nu}\,,
\label{eq:Yc:F}
\eeqn
where we took into account Eq.~\eq{eq:condition:1}.
Here $m^\mu_\cC(x) = \dot{x}^\mu_\cC/|\dot{x}_\cC|$ is tangent vector to the contour $\cC$,
where $\dot{x}^\mu_\cC \equiv \partial {\bar x}^\mu_\cC(\tau)/\partial
\tau$, and the contour $\cC$ is parameterized by the vector
function ${\bar x}^\mu_\cC$ of the variable $\tau$. Thus, if
\beqn
{\mathrm det}_{\perp} \bigl(m^\mu_\cC\, F^a_{\mu\nu}\bigr) \neq 0\,,
\label{eq:det}
\eeqn
then we have a hedgehog around the curve defined by the condition~\eq{eq:condition:1}. The determinant is taken
over indices $a$ and $\nu$, where the index $\nu$ is running in the 3D Lorentz subspace perpendicular to the
tangent vector $m^\mu_\cC$.

The hedgehog singularity has something to do with electric fields
rather than with the magnetic ones. Indeed in the case of a static
trajectory the matrix~\eq{eq:Yc:F} becomes a chromoelectric field,
$E^a_i \equiv F^a_{4i}$. Therefore it is difficult to associate
the discussed objects with magnetic monopoles despite the Abelian
monopole construction was explicitly used to define the objects.
The structure of the gluon fields of the object is dual to the
structure of the 't~Hooft-Polyakov monopoles which possesses a
hedgehog in the chromomagnetic sector.

One the other hand, it is clear that these hedgehogs should be
related in one or in another way to the confinement properties of
the system because their construction is done entirely in terms of
the Wilson loops (and, as it is well-known, the expectation value
of the Wilson loop variable is tightly related to the confinement
property of the Yang-Mills theory). The static hedgehogs should be
sensitive to the deconfinement phase transition since this
transition marks a change in the behavior of the electric
components of the gluon fields.

We expect that the physics of these objects is non-perturbative,
which can be uncovered, for example, by a numerical lattice
simulation which is one of the most powerful non-perturbative
methods. Unfortunately, a non-local definition of the hedgehogs
makes it difficult to locate this object in a given configuration
of the (lattice) gluon field. Moreover, a chance to find this
object directly in a lattice regularization is almost zero since
the Wilson loop is unlikely to be precisely center-valued. Loosely
speaking, the hedgehog goes through the meshes of the lattice and
it seems that the {\it individual} hedgehog is difficult to
locate. However, one can overcome this principal difficulty by
studying {\it statistical} rather than individual properties of
the hedgehogs. For example, in the SU(2) gauge theory one can
study a distribution $D_\cC(\omega) = \langle \delta(W_\cC -
\omega)\rangle$ of the trace of the Wilson loop $W_\cC$ at a
trajectory of a fixed shape $\cC$. The distribution is to be
evaluated at the gluon field ensemble. Then the density of the
center values of the Wilson loop $W_\cC$ can be obtained by an
extrapolation of the distribution $D_\cC(\omega)$ to the center
values, $w\to \pm 1$. It is clear that such limiting values are in
a general case finite quantity despite the center value $W_\cC =
\pm 1$ itself can not be reached exactly in the lattice
simulations. These distributions should provide information on the
density, correlation functions and other properties of these
objects.

\section{Examples}
\label{sec:examples}

\subsection{Vacuum configuration}

Consider the trivial vacuum configuration $A^a_\mu = 0$. In this case all untraced Wilson loops
are belonging to the center of the group, $W_\cC \equiv \bbbone$. However, the hedgehog
structure cannot obviously be found in the vicinity of any of these contours. Thus,
despite Condition 1, Eq.~\eq{eq:condition:1}, is fulfilled, Condition~2, Eq.~\eq{eq:condition:2}, is not:
the trivial configuration gives no singularities according to our criteria.

\subsection{BPS monopole}

The self-dual BPS monopole solution~\cite{ref:BPS} to the SU(2) Yang--Mills equation of motion is
\beqn
A^a_i & = & \frac{1}{g}\, f(r) \varepsilon_{iab} n^b\,,\qquad f(r) = \frac{1}{r} \Bigl(1 - \frac{r}{\sinh r}\Bigr)\,,
\label{eq:BPS:spatial}\\
A^a_4 & = & \frac{1}{g}\, h(r) \, n^a\,,\qquad \hspace{5mm} h(r) = \frac{1}{r} \Bigl(r \coth r -1 \Bigr)\,,
\label{eq:BPS:temporal}
\eeqn
where $n^a = x^a/|x|$, and $r \equiv |x|$ is assumed to be scaled by an arbitrary factor, $r \to r_0 \cdot r$,
to make a dimensionless quantity. The solution is static and self-dual.
The hedgehog configuration may only be present in the vicinity of the monopole, ${\mathbf x} = 0$.
Since the monopole is anyway static, let us consider the periodic boundary conditions:
the time direction is assumed to be compactified to a circle with the length $T$.

\begin{figure}[!thb]
\begin{center}
\includegraphics[scale=1.2,clip=true]{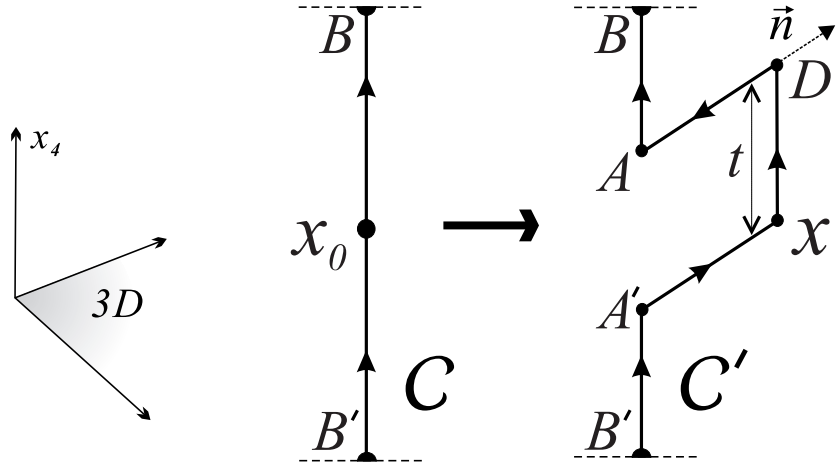}
\end{center}
\vspace{-4mm}
\caption{The contours used as probes for the BPS monopole.}
\label{fig:contour:rectangular}
\end{figure}

Consider an untraced Wilson loop which coincides with an untraced Polyakov loop,
\beqn
\cP_\cC(x) \equiv W_\cC(x) = W(x_0,B) \cdot W(B',x_0)\,, \qquad W(x,y) \equiv P \exp\Bigl\{i g \int_x^y \dd x_\mu\, A_\mu\Bigr\}\,,
\label{eq:W:static}
\eeqn
where integrations are taken along the straight contour $\cC$ parallel to the time direction, Figure~\ref{fig:contour:rectangular}.
Due to periodic boundary conditions the matrix $\cP_\cC(x)$ in
Eq.~\eq{eq:W:static} transforms as an adjoint operator. Using
Eq.~\eq{eq:BPS:temporal} one can easily see that the
condition~\eq{eq:condition:1} is fulfilled at the point ${\mathbf
x}=0$ as well as at the spheres $r = r_n$ with $T h(r_n) = 2 \pi
n$, $n\in \Z$. One can easily show that at the spheres $r=r_n$ due to the absence of the
isolated singularities the hedgehog condition~\eq{eq:condition:2} is not fulfilled.
Therefore below we consider the point ${\mathbf x}=0$ only.

We deform the contour $\cC \to \cC'$
as it is shown in Figure~\ref{fig:contour:rectangular}. Since $A_4({\mathbf x}=0)=0$
and the points $A$, $A'$, $B$ ad $B'$ are located at ${\mathbf x}=0$ line, then
$W(B',A')=W(A,B) = \bbbone$. Due to the structure of the spatial components of the gluon field~\eq{eq:BPS:spatial}
we have $W(A',x)=W(D,A) = \bbbone$. Finally, we find that the only nontrivial element of the path is $xD$,
and we get (taking $t\to 0$ and $r \to 0$):
\beqn
\cP_{\cC'}(x) & = & \bbbone \cdot \cos h(r) t/2 + i \sigma^a n^a \cdot \sin h(r) t/2 = \bbbone +
i \frac{t}{6} \cdot \sigma^a x^a + O(t^2,r^2)\,,\\
\Gamma_{\cC'}(x) & = & i \frac{t}{6} \cdot \sigma^a x^a + O(t^2,r^2)\,.
\eeqn
Thus, we obtain that condition~2, Eq.~\eq{eq:condition:2}, is
fulfilled at ${\mathbf x}=0$. The matrix $\cY_\cC$,
Eq.~\eq{eq:Y:variation}, is an infinitesimally small matrix
proportional to a unit matrix. The determinant of $\cY_\cC$ is
non--zero, thus we have clearly a non-degenerate case.
Therefore, for the self-dual BPS monopole configuration our
construction of the hedgehog gives the correct
location of the monopole center. Technically, our definition of the
gauge-invariant object, has identified the hedgehog-like structure of
the chromoelectric components of the self-dual BPS monopole.

\subsection{Polyakov Abelian gauge}

The Polyakov Abelian gauge is defined by the diagonalization of
(untraced) Polyakov loops $P_\cC(x)$, Eq.~\eq{eq:W:static}. The
isolated Abelian monopoles in this gauge are always static. The
Polyakov gauge has analytically been considered in
Refs.~\cite{ref:Yukawa,ref:Polyakov:gauge,ref:Jahn}. The monopole
positions are located by conditions~\eq{eq:X.eq.0} and
\eq{eq:X:prelim} in which the operator $X$ is identified with
$\cP_\cC(x)$. The Polyakov-loop variable can also be used to find
(static) monopole constituents in physically interesting
topologically non-trivial configurations~\cite{ref:Falk}. The
static BPS configuration in the Polyakov Abelian gauge corresponds
to an Abelian monopole. Thus, our recipe determines the monopole
position in the Polyakov gauge correctly if the background
configuration is the BPS monopole or the like. However, in a case
of a general configuration our construction and the definitions
for an Abelian monopole in the Polyakov Abelian gauge may give
different results.

\subsection{Dynamics at finite temperature}

At a finite temperature the Euclidean space-time is compactified
in one of the directions (it is called "imaginary time", or
"temperature" direction). As the temperature increases, the length
of the compactified direction becomes shorter and the gauge field
is forced to be static. This immediately implies, that our objects
-- located by
conditions~(\ref{eq:condition:1},\ref{eq:condition:2}) -- must
also become more and more static as temperature gets higher.
Indeed, if the object moves along a spatial direction at a point
$x$, then the gauge field must evolve in the temporal direction to
support the hedgehog structure~\eq{eq:condition:2}. However, the
evolution of the fields in the temporal direction -- and,
therefore, the motion of the object in a spatial direction -- is
suppressed at high temperatures.

Moreover, at low temperatures ($i.e.$, in the confinement phase)
the distribution of the values of the Polyakov lines is peaked
around $(1/2)\, {\mathrm{Tr}} \cP(x) = 0$ value, while at high
temperatures the Polyakov loops tend to be concentrated near
$(1/2)\, {\mathrm{Tr}} \cP(x) = \pm 1$ values supporting condition
1, Eq.~\eq{eq:condition:1}. The described behavior of the Polyakov
lines indicates that as the temperature increases the density of
static objects becomes higher and higher compared to the density
of the spatial currents. Similar property is observed for the
Abelian monopoles in lattice simulations~\cite{ref:finite:T}.

\subsection{Lower dimensions}

In lower dimensions the similar object can also be formally
defined by condition 1. However, the hedgehog
condition~\eq{eq:condition:2} can not be fulfilled since there is
no hedgehog structure (in a monopole sense) around the trajectory
$\cC$ in two spatial dimensions. This consideration means that the
hedgehog structure around the object should be of the
chromoelectric nature.

\section{Discussion and Conclusion}

We propose a gauge--invariant definition of a hedgehog-like object
in non-Abelian gauge models. The trajectory of this object is a
closed curve defined by the requirement for the (untraced) Wilson
loop to take its value in the center of the color group. This
definition shares similarities with 't~Hooft definition of an
Abelian monopole and locates correctly the trajectory of the
self-dual BPS monopole. One the other hand, the hedgehog-like
behavior is encoded in the chromoelectric components of the gluon
field, implying that the hedgehogs should be sensitive to the
finite temperature phase transition. We provided arguments that
the density of the static hedgehogs should increase with the rise
of temperature.

It is interesting to check the properties of these objects on the
lattice despite the lattice definition of an individual hedgehog
in a configuration of the gluon fields is somewhat obscure. We
have proposed a method to implement our construction by studying
statistical rather than individual properties of the hedgehogs
with the help of distributions of the Wilson loops and/or their
eigenvalues at trajectories of fixed shapes.

\begin{acknowledgments}
The author is supported by grants RFBR 04-02-16079, MK-4019.2004.2
and by JSPS Grant-in-Aid for Scientific Research (B) No.15340073.
The author is grateful to F.V.Gubarev, M.I.Polikarpov, T.Suzuki, V.I.Zakharov
and the to members of the ITEP Lattice group for useful
discussions. The author is thankful to the members of Institute
for Theoretical Physics of Kanazawa University for the kind
hospitality and stimulating environment.
\end{acknowledgments}

\end{document}